\documentclass[aps,floatfix,twocolumn,amsmath]{revtex4}
\usepackage{graphicx}
\newcommand{\be}{\begin{equation}}
\newcommand{\ee}{\end{equation}}
\newcommand{\bea}{\begin{eqnarray}}
\newcommand{\eea}{\end{eqnarray}}

\begin{document}

% Use the \preprint command to place your local institutional report
% number in the upper righthand corner of the title page in preprint mode.
% Multiple \preprint commands are allowed.
% Use the 'preprintnumbers' class option to override journal defaults
% to display numbers if necessary
%\preprint{}

%Title of paper
\title{Acceleration of the Universe via $f(R)$ Gravities \\and \\The Stability of Extra Dimensions}

% repeat the \author .. \affiliation  etc. as needed
% \email, \thanks, \homepage, \altaffiliation all apply to the current
% author. Explanatory text should go in the []'s, actual e-mail
% address or url should go in the {}'s for \email and \homepage.
% Please use the appropriate macro foreach each type of information

% \affiliation command applies to all authors since the last
% \affiliation command. The \affiliation command should follow the
% other information
% \affiliation can be followed by \email, \homepage, \thanks as well.
\author{Tongu\c{c} Rador}
\email[]{tonguc.rador@boun.edu.tr}

%\homepage[]{Your web page}
%\thanks{}
%\altaffiliation{}
\affiliation{Bo\~{g}azi\c{c}i University Department of Physics \\ 34342 Bebek, \.{I}stanbul, Turkey}

%Collaboration name if desired (requires use of superscriptaddress
%option in \documentclass). \noaffiliation is required (may also be
%used with the \author command).
%\collaboration can be followed by \email, \homepage, \thanks as well.
%\collaboration{}
%\noaffiliation

\date{\today}

\begin{abstract}
% insert abstract here
We discuss the possibility of obtaining the present acceleration of the universe via $f(R)$ gravity theories which recently attracted much attention. It is known that $f(R)$ theories generally have room for this. In this work we stress that the requirement for the
stabilization of extra dimensions is naturally incorporated in such a generalization of Einstein gravity under rather orthodox assumptions. We have restricted our study to pure $f(R)$ gravity without additional matter sources partially in view of the fact that
if the acceleration is to continue indefinitely any ordinary matter term is to redshift to irrelevancy and mostly for economy in understanding. The general conditions we find is that the manifold of the extra dimensional space is to have negative internal curvature and that the Ricci scalar of the full space-time manifold is also negative. The positive curvature case for extra dimensional manifold is actually the most generic case. However this necessitates a fine tuning between the Hubble constant and the size of extra dimensions in the absence of matter sources.
\end{abstract}

% insert suggested PACS numbers in braces on next line
%\pacs{02.50.-r, 05.40-a}
% insert suggested keywords - APS authors don't need to do this
%\keywords{}

%\maketitle must follow title, authors, abstract, \pacs, and \keywords
\maketitle
\section{Introduction}

Recent experiments show that the observed universe expand in an accelerating fashion. The most economical way of explaining this in 4-dimensional physics is via a cosmological constant which would also imply that the acceleration will be persistent in the future. In fact if acceleration is not something that happens randomly this seems to be the most plausible approach. Of course the cosmic scandal \cite{z0} (a term coined to stress the fact that the acceleration setting in in the recent past requires considerable fine tuning) is possibly the hardest to deal with. One way out of this dilemma is to propose that acceleration is something that happens every now and then \footnote{We do not mean a small period of acceleration that happens once. This can be obtained very easily and arbitrarily in a given theory with contrived initial conditions. We advocate that this is the least aesthetic way to deal with the cosmological scandal problem} or at least something that happens with considerable probability.

Another issue with the acceleration of the observed space is the fact that assuring the stability of extra dimensions (if they exist) is rather difficult. The existence of extra dimensions is a necessity for string theory and their stability can be achieved rather generally for power-law, non-inflating solutions for the scale factor of the observed space. For example, within string gas or brane gas cosmology \cite{a0} -\cite{a31} the stabilization can be achieved via the interplay between the momentum and winding modes of strings and branes. But such sources in general yield solutions for the observed dimensions similar to presureless dust (a necessary fact since these objects are confined to the extra dimensional space and have no pressure along observed dimensions).

In this work we follow a phenomenological approach. We study the conditions for which a solution with stable extra dimensions and inflating observed dimensions exist. We show that within Einstein gravity (which is supposed to be part of the low energy limit of string theory assuming a constant dilaton) such a solution necessitates a rather contrived source with pressure coefficient of -2 along internal dimensions. Such a source would be hard to accommodate. Motivated by this situation we discuss the possibilities for a generalization of Einstein gravity by including higher order terms of the scalar curvature to the action, namely $f(R)$ gravities. These theories attracted much attention recently \cite{fr2}-\cite{fr21} (see also \cite{fr0}-\cite{fr1} for earlier works) but these studies were generally in 4 dimensions  and therefor are free from the problem of the stability of extra dimensions (see however \cite{frd0}-\cite{frd1} for recent d-dimensional cases which are somewhat orthogonal to our approach and the note added at the end of this manuscript). 

We analyse the behaviour of such theories with the inclusion of extra dimensions. The general conditions we find is that in the absence of sources, the manifold of the extra dimensional space is to have negative internal curvature and that the Ricci scalar of the full space-time manifold is also negative. The omission of sources can be advocated in the following way: if acceleration is to happen indefinitely then any ordinary source term will have to redshift and become irrelevant in the future. Furthermore, in $f(R)$ theories a cosmological constant which is the source that would not redshift is not necessary to obtain inflation. So from these
perspectives we have omitted any sources to see whether one can obtain inflating observed dimensions with stable internal space.

The positive curvature case for extra dimensional manifold is actually the most generic case. However this necessitates a fine tuning between the Hubble constant and the size of extra dimensions in the absence of matter sources.

\section{Pure Einstein Gravity}

In a cosmological scenario one can take

\be
ds^{2}=-dt^{2}+e^{B(t)}\sum_{i=1}^{m}dx_{i}^{2}+e^{C(t)}\sum_{a=1}^{p}dy_{a}^{2}\;\;,
\ee

\noindent where $B$ ($C$) represents the scale factors of the $m$ ($p$) dimensional observed (extra) space. The dimensionality of the full space-time is $d\equiv 1+m+p$. Following the experimental status we assume that the observed dimensions are flat and non-compact. We also allow for internal curvature for the manifold of extra dimensions. We are interested in the following solution to account for the observed acceleration of the universe and the
stability of the extra dimensions,

\begin{subequations}
\bea{\label{eq:anz}}
B(t)&=&Ht+B_{o}\;\;,\\
C(t)&=&C_{o}\;\;.
\eea
\end{subequations}

The most economical way to incorporate acceleration in the observed dimensions is via a cosmological constant but this cannot stabilize extra dimensions \footnote{One can try to do it via a positive curvature along the extra dimensions but this is analogous to the Einstein static universe and is not a stable equilibrium point}. Possibly the next best thing is to assume a component for the energy density such that one has pressure coefficient $\omega=-1$ along the observed dimensions so that it never redshifts like a cosmological constant,

\be{\label{eq:src}}
\rho=\rho_{o}\exp\left[-(1-1)mB-(1+\nu)pC\right]\;\;.
\ee

\noindent One can easily check that this source, having the form of a hydrodynamical fluid, yields a conserved energy momentum tensor. This will result in the following equations for the scale factors

\begin{subequations}
\bea
\ddot{B}+\dot{k}\dot{B}&=&\frac{2-(1+\nu)p}{d-2}\rho\;,\\
\ddot{C}+\dot{k}\dot{C}&=&\frac{(m+1)+(m-1)\nu}{d-2}\rho-2\kappa\;e^{-2C}\;,\\
\dot{k}^{2}&=&m\dot{B}^{2}+p\dot{C}^{2}+2\rho-2\kappa e^{-2C}\;, \\
k&\equiv&mB+pC\;.
\eea
\end{subequations}

\noindent Here $\kappa$ represents the internal curvature along the extra dimensions and one has $\kappa>0$, $\kappa<0$ of $\kappa=0$ \footnote{Note that we are using units in which $\kappa$ has dimensions.}. Stability can only be achieved for $\rho_{o}>0$ and $\kappa<0$. Thus the manifold of extra dimensions must have negative curvature. Negative internal curvature have been a constant point of attention in the past for cosmology, brane worlds and other concepts involving extra dimensions. See for examples \cite{trod1}-\cite{trod4} and the references therein.

Unfortunately this is not the only condition one must also have

\be
\nu\leq-\frac{m+1}{m-1}\;.
\ee

\noindent The equality is allowed for $\kappa=0$ where one looses the information on where $C_{o}$ stabilizes

\be
\frac{(m+1)+(m-1)\nu}{d-2}\rho(C_{o})-2\kappa\;e^{-2C_{o}}=0\;.\\
\ee

\noindent So $\kappa=0$ does not stabilize extra dimensions but it is compatible with an already stabilized radion.

The problem we immediately see is that for $m=3$ one gets the following for $\nu$

\be{\label{eq:crazy}}
\nu\leq-2\;\;\;!
\ee

If one cannot come up with a model that predicts this pressure coefficient the following possibilities are natural

\begin{itemize}
\item{} There are other possible sources with $\omega=-1$ and different $\nu$'s.
\item{} $\omega\neq-1$ by a large deviation.
\item{} Extra dimensions are not stable.
\item{} Possible $\rho_{o}<0$ sources enter the game.
\item{} Einstein gravity is not the complete low energy limit of a fundamental theory.
\end{itemize}

\noindent Of course one can also have a combination of any of these possibilities. 

Unfortunately adding other sources of the form (\ref{eq:src}) with different pressure coefficients will not change anything: on of them will still have to satisfy (\ref{eq:crazy}). The limitations on the motion of extra dimensions come from the very stringent bounds on the time variation of the fundamental constants, so we assume they are not evolving. The fits on $\omega$ strongly favor $\omega=-1$ with only a small room for variance. There are various approaches with negative energy density sources, such as a rolling tachyon but these approaches are as bit as unique (or maybe contrived) solutions as setting $\nu\leq-2$. This seem to leave us with the fourth choice.

%\section{Dilaton Theories}
%

\section{$f(R)$ Gravities}

Higher derivative theories generally suffer from the Ostrogradski  instability \cite{os} (and \cite{os2} for a nice review) but $f(R)$ theories evade this

\be
S=\int dx^{d}\sqrt{-g}f(R)\;,
\ee

\noindent with $f$ being an algebraic function of the curvature scalar. We do not include any matter terms at this point to see if $f(R)$ is sufficient for our purposes \footnote{A cosmological constant is still a possibility as part of $f(R)$ and any other source will have to redshift anyways. So assuming the present acceleration of the universe will go on indefinitely it is justifiable to ignore other sources. Except maybe sources of the form (\ref{eq:src}) can be studied but this does not do justice to the economical approach to explain the current situation.}. This yields the following equations of motion

\be
f' R_{\mu\nu}-\frac{1}{2}g_{\mu\nu}f+\left[g_{\mu\nu}\nabla^{2}-\nabla_{\mu}\nabla_{\nu}\right]f'=0\;.
\ee

\noindent where prime denote derivative with respect to $R$. Here as usual, one can trade the $0-0$ equation for the trace equation

\be
f'(R)R-\frac{d}{2}f(R)+(d-1)\nabla^{2}f'(R)=0\;.
\ee

As before we are interested in stable solutions of the form in (\ref{eq:anz}). This means that $R=R_{o}$ but not $R_{\mu\nu}=R_{o}g_{\mu\nu}/d$ since we only allow internal curvature along extra dimensions. A necessary condition for stability can be obtained from the trace equation by expanding around $R_{o}$ and linearizing \footnote{This condition is that the scalar degree of freedom that is excited in $f(R)$ theories (namely $f'(R)$) should not have negative mass squared. For a general linear stability analysis one will have to introduce small perturbations for the scale factors as $\delta B$ and $\delta C$. This is not a difficult task if we remember that the
solution around which we are expanding is not a power-law solution and the perturbation equations will generally be linear differential equations with constant coefficients and the analysis is rudimentary. We do not explicitly expose it here.},

\be{\label{eq:stab2}}
\frac{d-2}{2}\frac{f'(R_{o})}{f''(R_{o})} > R_{o}\;\;.
\ee

We therefor have to satisfy the following along with the stability condition

\begin{subequations}
\bea
f'(R_{o})R_{o}^{\hat{i}\hat{i}}=\frac{1}{2}f(R_{o})\;,\\
f'(R_{o})R_{o}^{\hat{a}\hat{a}}=\frac{1}{2}f(R_{o})\;,\\
f'(R_{o})R_{o}=\frac{d}{2}f(R_{o})\;.
\eea
\end{subequations}

\noindent Here $R_{o}^{\hat{i}\hat{i}}$ and $R_{o}^{\hat{a}\hat{a}}$ represent the elements, in the orthonormal basis, of the Ricci tensor along the observed and extra dimensions respectively. For the solution we are after they are given by,

\begin{subequations}
\bea
R_{o}^{\hat{i}\hat{i}}&=&mH^{2}\;\;,\\
R_{o}^{\hat{a}\hat{a}}&=&2\kappa\;e^{-2C_{o}}\;\;,\\
R_{o}&=&m(m+1)H^{2}+2p\kappa e^{-2C_{o}}\;\;.
\eea
\end{subequations}

A quite general conclusion is that (again in the absence of sources) if $f(R_{o})\neq0$ and $f'(R_{o})\neq0$ one gets

\be
mH^{2}=2\kappa e^{-2C_{o}}\;\;.
\ee

\noindent which would mandate $\kappa>0$ and an fine tuning between $H$ and $C_{o}$. Even worse it will predict a value for $H$ which would be $O(1)$ in Plank units if extra dimensions are stabilized at that scale.  Stating the same from the opposite direction one will have to have very large extra dimensions if one gets $H$ from experimental data. So this cannot be phenomenologically viable 
\footnote{Of course if $\kappa=0$ we get $H=0$ which means that the full manifold is $R\times R^{m}\times T^{p}$ this solution is not necessarily stable in Einstein theory and would evolve into
Kasner solutions under arbitrary perturbations. One can also show that the situation is similar for $f(R)$ theories, that is the space $R\times R^{m}\times T^{p}$ is not linearly stable under arbitrary perturbations of the scale factors.}.

A simple way out of this fine tuning bind without adding sources and still have $H\neq0$ is to demand $f(R_{o})=f'(R_{o})=0$ with $f''(R_{o})\neq0$. The stability condition in this case would mean

\be\label{eq:stz}
R_{o}<0\;\;.
\ee

\noindent which can only be satisfied if $\kappa<0$ and $\left|2p\kappa e^{-2C_{o}}\right|>m(m+1)H^{2}$ which is a less restrictive case. It is interesting that the condition on $\kappa$ survives the generalization of Einstein theory to $f(R)$ models.

Recently theories involving negative powers of $R$  have been studied \cite{fr2}-\cite{fr10}. But these models are generally
in trouble with solar system data and other considerations. For example $1/R$ theories predict a force a million times larger between Andromeda and Milky Way galaxies ! \cite{os2}.

We therefor see that $f(R)$ theories quite generally have room for accelerating observed dimensions while keeping extra dimensions stable. Of course this represents a rather large class of theories. To bound this large class one could for example request that $R\times R^{m}\times T^{p}$ (where $T^{p}$ represents p-dimensional torus) or any other case with $R_{o}=0$ (stable or not) be a solution. This can be justified with the pretext that quantum field theory works in flat and static space-times which constitutes a subset of $R_{o}=0$ solutions. This will possibly put away theories with negative powers of $R$ \footnote{See however \cite{fr18}-\cite{fr19} for some counter examples to this idea where the space-time dimensionality is 4 and Minkowski space can in certain cases be a solution even with negative powers. Since if there are extra dimensions the full manifold is not Minkowski we believe the argument in the text about $R_{o}=0$ solutions still has some weight.}. Furthermore one can also try to get rid of the finely tuned case by requesting that there are no $R_{o}>0$ solutions to the equilibrium conditions which means that $f(R_{o})=f'(R_{o})=0$ and $f''(R_{o})\neq0$. These considerations narrows down the possible theories somewhat. But in any case there are still infinitely many possibilities.

We would however like to discuss a class of theories which was not discussed yet in the literature. Namely the ones that involve infinitely many powers but yet have a $f(R)$ bound from above and below. Consider for instance the following

\be
f(R)=\bar{R}\left[\sin(\frac{R}{\bar{R}})+\sin^{2}(\frac{R}{\bar{R}})\right]
\ee 

One can show that the following is a stable point

\be{\label{eq:fin}}
R_{o}=m(m+1)H^{2}+2p\kappa e^{-2C_{o}}=-\frac{4n+1}{2}\pi\bar{R}\;.
\ee

\noindent with $n$ being a positive semi-definite integer. The solutions with $R_{o}\geq0$ are not in accord with the stability condition. As discussed this will happen only for $\kappa<0$. If one uses normalized coordinates such that $\kappa=-\alpha\bar{R}$ with $\alpha$ dimensionless this would imply that allowed $C_{o}$ values in this solution will be discrete. Seeing it from another perspective is fixing $C_{o}$ which would imply that a very large $H$ can be analogous to a small $H$ as well. With such an infinite number of possible solutions one can hope for an understanding of the present acceleration of the
universe as well as  the cosmic scandal problem.

So theories with $f(R_{o})=f'(R_{o})=0$ and $f''(R_{o})\neq0$ along with the condition $R_{o}<0$ are accommodating solutions of the form (\ref{eq:fin}). 

A rather interesting consideration is that $H=0$ is not necessarily stable for it yields the following solutions for a linear perturbation 

\begin{subequations}
\bea
\delta B=\delta B_{o}+\delta H t\;\;,\\
\delta C=\delta C_{1}\cos\left(\omega t\right)+\delta C_{2}\sin\left(\omega t\right)\;\;.
\eea
\end{subequations}

\noindent with $\omega=\sqrt{2|\kappa|e^{-2C_{o}}}$. So a solution of the form $R\times R^{m}\times M_{p}$ with $M_{p}$ having negative internal curvature will evolve into an acceleration for the expanding large dimensions under small perturbations. Possibly the smallness of the observed $H$ can also be accommodated this way: a small oscillation of the size of extra dimensions will trigger acceleration of the observed space if one initially is in the $R\times R^{m}\times M_{p}$ state.

\section{Discussion}

We have discussed the generalization of Einstein gravity to $f(R)$ theories including higher powers of the Ricci scalar. Our motivation for doing so is that within Pure Einstein gravity it is rather contrived to have accelerating observed dimensions and stable extra dimensions. Another reason for picking only $f(R)$ gravities is that they are generically free of the Ostrogradski instability. We have shown that the sought after solution can be generally accommodated within a subclass of $f(R)$ theories as described in the text. This subclass has the solution as a stable point at least in the linear sense. The requirement we find is that the manifold of the extra dimensional space must have negative curvature and the the full Ricci scalar must also be negative at these stable points. We also show that there are other solutions where the manifold of extra dimensions have positive curvature but these cases requires fine tuning between the present size of extra dimensions and the observed Hubble constant. In view of this fine tuning we did not further discuss these cases.

We have omitted theories with higher derivatives of the Ricci scalar again in view of the fact that they would suffer from Ostrogradski instability. Recently theories involving infinitely many derivatives were suggested as an alternative for the early inflationary paradigm \cite{x1}-\cite{x4}. One can argue that these theories might be free of the mentioned instability because one cannot isolate the highest order momenta. One can possibly accommodate acceleration for the observed space and stable extra dimensions within such a framework. But we did not discussed it here.

We have also omitted dilaton type theories without extra powers of the Ricci scalar. For example string theory predicts the following action as a low energy theory

\be
S=\int\;\frac{dx^{d}}{G_{4+d}}\sqrt{-g}e^{-2\phi}\left[R+4(\nabla\phi)^{2}+\kappa^{2}\mathcal{L}\right]\;.
\ee

It is evident that if the dilaton stabilizes via some mechanism from $\mathcal{L}$ one will have Einstein theory and we would still face the need for a source with $\nu\leq-2$. The rather general meaning for this is that to avoid such a pressure coefficient one will have to allow for a changing dilaton and this in turn means that the 4 dimensional gravitational constant is time-dependent.  Any other dilaton theory with a non-minimal coupling will have this general feature. This could in fact be what is happening in nature but do not advocate this possibility in this work.

One possible generalization of the dilaton theories is to endow them with higher order terms in the Ricci scalar. This is of course a possibility. For a constant dilaton one would be duplicating our discussion. Therefor any other approach which would be distinct would involve a rolling dilaton. We did not discuss this case as it would complicate the main emphasis of this work.

We have also omitted any matter sources in this work and this was in fact a key argument in stating that the positive curvature solutions for the internal space implies a fine tuning between the Hubble constant and the size of extra dimensions. So one possible
extension of this work is to extend the study such that one includes sources. An idea is to have sources without redshift as the one
presented in the first section. This type of sources are the only type which resembles a cosmological constant from a four dimensional point of view and yet, unlike cosmological constant, cannot be accommodated in a pure $f(R)$ approach. This extension is left for future study.

\subsection{Note Added}

The conclusions of this paper are in accord with the earlier works of Zhuk et. al. \cite{zhuk1}-\cite{zhuk5} considering also d-dimensional models with an emphasis on the stability of extra dimensions. Their analysis relied on finding the set of conformal transformations which would transform a d-dimensional $f(R)$ gravity to a 4-dimensional ordinary Einstein theory with two scalars. One of these scalars is the radion field and the other is the field related to the $f'(R)$ excitation in the d-dimensional case. I thank Prof. Zhuk for his comments.

\end{document}